# Towards efficient random metasurfaces

Hadiseh Nasari, Matthieu Dupré, and Boubacar Kanté*

Department of Electrical and Computer Engineering, University of California San Diego, La Jolla, California, 92093-0407, USA.

*Corresponding author: bkante@ucsd.edu

## Abstract

Random media introduce large degrees of freedom in device design and can thus address challenges in manipulating optical waves. Wave shaping with metasurfaces has mainly utilized periodic or quasi-periodic grids, and, the potential of random arrangement of particles for devices has only come under investigation recently. The main difficulty in pursuing random metasurfaces is the identification of the degrees of freedom that optimize their efficiencies and functions. They can also encode information using the statistics of particles distribution. We propose a phase-map that accounts for the statistical nature of random media. The method takes into account effects of random near-field couplings that introduce phase errors by affecting the phase shift of elements. The proposed approach increases the efficiency of our random metasurface devices by up to ~20%. This work paves the way towards the efficient design of random metasurfaces with potential applications in highly secure optical cryptography and information encoding.

## Introduction

Random media, by introducing unique features that are absent in their periodic counterpart have been intensively investigated during the last few decades [1-9]. Addressing challenges of wave scattering in random media could be precious to overcome and manage the intrinsic randomness of biological tissues for imaging and for the treatment of cancer tumors deep inside tissues [10]. It has also been shown that multiple scattering of light in a disordered medium made of randomly distributed nanoparticles can improve imaging performance by overcoming the diffraction limit of conventional imaging approaches and also extending its field of view [11-14]. The insensitivity to the polarization of light, even for anisotropic and asymmetric particles, and, the absence of spurious diffraction orders present in periodic structures with periodicity comparable to the wavelength, are other advantages introduced by random media [15, 16]. The potentially simpler fabrication process of random structures, not requiring precise positions, enables self-assembly approaches [17]. Random or disordered metasurfaces, compared to more conventional three-dimensional disordered media with thicknesses much larger than the operating wavelength, are easier to implement and exhibit lower losses providing a promising platform for light manipulation [18-20]. While most metasurfaces employed for wave shaping are patterned in an ordered or quasi-ordered manner, it has been shown that a random platform can be engineered to create desired wave fronts.

In wave shaping with phase controlled metasurfaces, the incident light should ideally undergo a specific but transversally continuous phase change upon transmission or reflection. However, in reality, this spatial phase profile on the metasurface area is discretized by sampling and then implemented in a local manner by particles at specific positions with the appropriate geometry, orientation, and material composition [21-24]. Regardless of the sampling approach, particles should provide the correct phase at the selected sampling point on the surface so that the device may have the highest efficiency. The abacus providing the geometry of a particle for a desired phase response is denoted phase-map in the paper (see Fig.1). The question is then, what is the best phase-map for a specific sampling geometry in order to maximize the efficiency of metasurface devices? In particular, what is the best phase-map when the phase of the metasurface is randomly sampled? This question is all the more important and difficult that particles on the metasurface are surrounded by other particles leading to random near-field coupling. In Fig. 1, the black particle is surrounded by purple particles that affect its phase and near-field coupling can go beyond the nearest neighbors. The selection of the geometry of the black particle (the prediction of its actual phase) should at least account for the nearest neighbors especially for dense metasurfaces. In this paper, we propose, for the first time to the best of our knowledge, a phase-map for random metasurfaces that accounts for the statistical nature of random media and show that it improves the efficiency of meta-devices.

## Simulation Results and Discussion

To illustrate our approach, we consider the metasurface presented in Fig. 1(a), composed of titanium dioxide (TiO$_2$) cylindrical elements randomly distributed on a surface, and implementing a lens in reflection. The method can be generalized to other particle shapes and materials composition[25,26]. In this reflection geometry, a single Mie resonance is sufficient to achieve the $2\pi$ phase control needed to implement a low-loss metalens. Random meta-devices working in transmission mode can be similarly designed but two resonances will be needed to achieve the required phase change [27,28]. Our lens requires a specific phase profile at each position on the metasurface given by [29,30]

$$\varphi(\mathcal{R}) = 2\pi\left[\sqrt{\mathcal{R}^2 + f^2} - f\right]/\lambda) \qquad (1)$$

Where $f$ is the focal length, $\lambda$ is the wavelength of the incident light and $\mathcal{R} = \sqrt{x^2 + y^2}$ is the distance from an arbitrary position (x, y) on the lens to the center of the metalens.

The random metasurface presented in Fig. 1(a) is generated numerically using a method previously proposed [15]. For a chosen position on the random metasurface, the geometry of the particle giving the required phase (Fig. 1(b)), is chosen using the periodic phase-map presented in Fig. 1(c).

Let's recall the basics of periodic sampling with 2D square grids. To achieve $2\pi$ phase shift, the radius of the cylindrical element of Fig. 1(b) is tuned over a suitable range. This radius tuning affects a resonant mode of the element and the phase shift made by the element at a given frequency. To calculate the phase-map, in periodically sampled metasurfaces, a large array of basic elements arranged in a periodic configuration is usually used. Numerical simulations are conducted on a unit-cell shown in Fig. 1(b) to calculate the phase. This method, known as the unit-cell method in the literature has been successful for most metasurfaces. The resulting phase-map, that we call periodic phase-map, is plotted in Fig. 1(c). In the actual metalens, neighboring elements are of different size, which affects the phase shift that the particle effectively provides, resulting in phase errors and thus a lower efficiency. Indeed, in presence of near-field coupling, the radius of cylinders should be modified to compensate for this phase error. A local phase approach using optimization algorithms can resolve this concern as reported in [31]. In the case of random sampling, the problem is exacerbated and the periodic phase-map is far from being optimal, and, it may fail to determine the radius of the elements as both the size, the position, and orientation for anisotropic particles, of neighbors contribute to phase errors.

Figure 1(d) shows the phase profile of a metalens with random sampling but using elements whose dimensions are chosen from a periodic phase-map (Fig. 1(c)) with the periodicity of 510 nm. At first glance, it seems that the local-phase method and optimization approaches can also be used for this case, however, the many sources of errors (listed above) would make such optimization computationally very expensive for large metasurfaces.

To bring an insight on how randomness in the position of elements affects the efficiency of the metalens, we first investigated a jittered random array. In this type of random metalens, elements are displaced from their periodic positions according to $x_r = x_p + \mu$ and $y_r = y_p + \mu$ where $(x_p, y_p)$ and $(x_r, y_r)$ are the position of elements in the periodic and random arrays respectively and $\mu \in (-\mu_{max}, \mu_{max})$ is a random value. Figure 2 illustrates how adding randomness up to 80 nm in the position of elements in the periodic metalens with a grid periodicity of 510 nm, and, updating the radius of elements using the periodic phase-map of Fig. 1(c), reduces the efficiency by almost half (Fig. 2(a)) even though the metasurfaces still successfully focus light (Fig. 2(b)).

For fully random sampling of an ideal phase profile with N cylindrical particles, the required phase shift φ at two symmetric positions $c_1$ and $c_2$ with respect to the center of the lens would be provided by cylinders of dissimilar radius $r_1$ and $r_2$ because of the environment (i.e. neighboring elements) around the two position is in general different. The question then is: What is the ensemble of configuration (phase-map) that can provide $2\pi$ phase shift at the design wavelength while accounting for the complex environment of particles in random metasurfaces? We propose an approach that can reduce the computational cost of simulations and facilitate the design via random phase-maps. The random phase-maps are for random metasurfaces what periodic phase-maps, widely used currently, are for metasurfaces using periodic grids. They are constructed as follows.

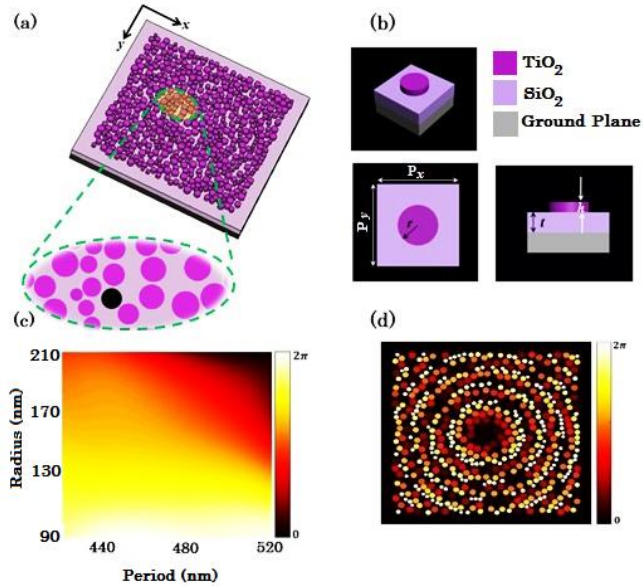

**Fig. 1.** (a) Schematic of the random metalens. (b) Unit-cell for calculating the periodic phase-map. (c) Periodic phase-map with the thickness of $SiO_2$ spacer t= 350 nm and the height of $TiO_2$ h= 200 nm. The index of $SiO_2$ and $TiO_2$ are 1.53 and 2.49 respectively. (d) Random sampling of the phase profile of the lens using the periodic phase-map in panel (c) for a periodicity of Px=Py=510 nm. The lens operates at $\lambda = 800\ nm$ and has a focal length of $f = 2\ \mu m$.

Starting with the periodic phase-map, the particle of largest size is randomly distributed on a surface at the highest density. The fully random array generated by this approach is called totally random array. This metasurface, made of particles of the same size, is generated using MATLAB, while preventing overlaps of particles [32]. The method consists in selecting a random position for the first particle and then a position for the next particles in a sequential manner while avoiding overlaps with previously placed particles until the maximum density is reached (see Fig. 3(a)).

Keeping the position of particles in the previous array, the radius is varied (reduced) resulting in an ensemble of random metasurfaces in which the only difference is the radius of particles (see Fig. 3(a) and 3(b)).

The phase shift provided by the previous metasurfaces is numerically computed using a supercomputing center. Unlike periodic phase-maps, each radius now corresponds to a random array, but arrays have the same randomness.

The three previous steps are repeated several times for different realization of randomness.

For each radius, the phase of various realizations of randomness is averaged and the standard deviation of the phase is calculated.

The previous five steps are repeated for larger array sizes (larger number of particles) until convergence in the average and standard deviation of the phase is attained.

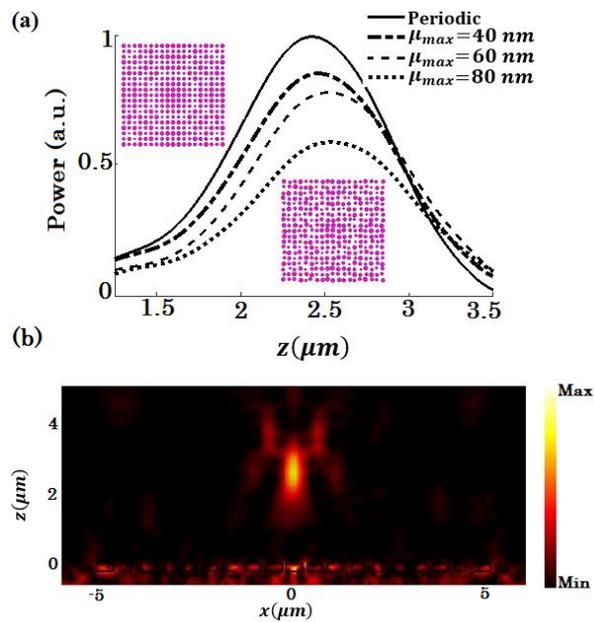

**Fig. 2.** Comparison of the focusing power of periodic and disordered metalens made of jittered random arrays for different strength of disorder. (b) Power profile in the x-z plane for the jittered random metalens with $\mu_{max} = 40\ nm$. The size of the computed metalens is about ~$14\lambda$ by $14\lambda$ area.

Figures 3(a) and 3(b) present the array of largest and smallest radius for one realization of randomness used to calculate the phase shift via the proposed algorithm. As seen, for one realization of randomness, the filling factor defined as the fraction of the area of the metalens occupied by elements decreases upon decreasing the radius of elements but the number and position of elements are kept constant.

Results for the random phase-map calculated by this approach and the standard deviation of the phase for different realization of random metasurfaces are presented in Fig. 3(c) and 3(d) respectively. The simulations are conducted using the time domain solver of the commercially available software CST. The convergence of numerical simulations is verified by decreasing the maximum size of the mesh unit until similar results are obtained between consecutive iterations and by ensuring that energy in the computational domain decreases continuously after excitation. The maximum difference between the random and periodic phase-maps is about $50^0$.

It is worth noting that in the calculation of the random phase-map, $2\pi$ phase may not be provided by an arbitrary number of particles or in an equivalent statement by all filling factors. Actually, there is a minimum filling factor above which $2\pi$ phase shift can be obtained. Here, we have achieved $2\pi$ phase shift by increasing the filling factor for cylinders of largest size to 59%. Metalenses are therefore configured at the highest possible density (filling factor).

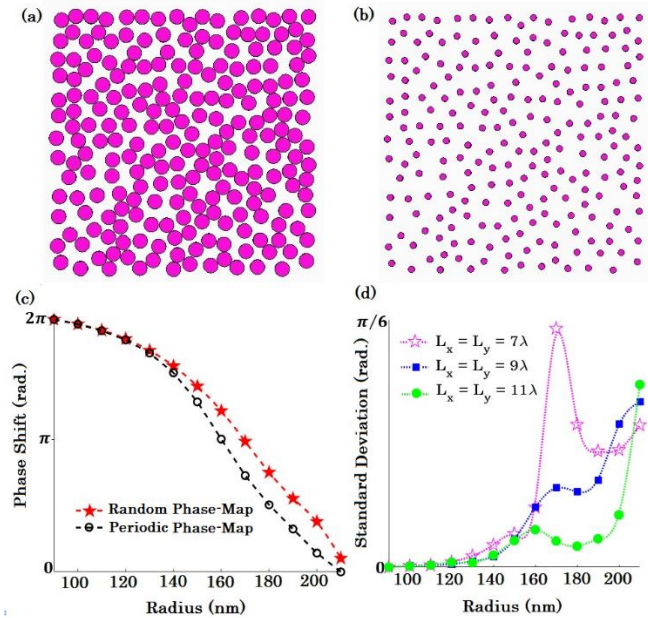

**Fig. 3.** Generated random arrays ($L_x$ by $L_y$) to calculate the phase shift of elements in a random planar arrangement for two different radii of cylindrical elements (a) 210 nm and (b) 90 nm. (c) Random phase-map obtained by averaging over several calculations of phase shift made by elements in various random arrangements (different realization of randomness) in comparison with the periodic phase-map (d) standard deviation of the calculated phases showing a reduction in divergence of the result when the area (number of elements) increases.

The random-phase-map presented in Fig. 3(c) shows that the phase shift obtained differs from the periodic phase-map for a larger radius of the particles, i.e., when near-field coupling is present. The proposed random phase-map, because it keeps constant the position of random particles for each randomness realization, does not account well for near-field coupling in the metasurface for particles of small radius (see Fig. 3 (b)).

Using the proposed random phase-map, Fig. 4 shows that the average focusing efficiency of the random metalens increases by ~20% compared to the random metalens implemented using the periodic phase-map [15]. The presented results are obtained by averaging over fifteen realizations of randomness.

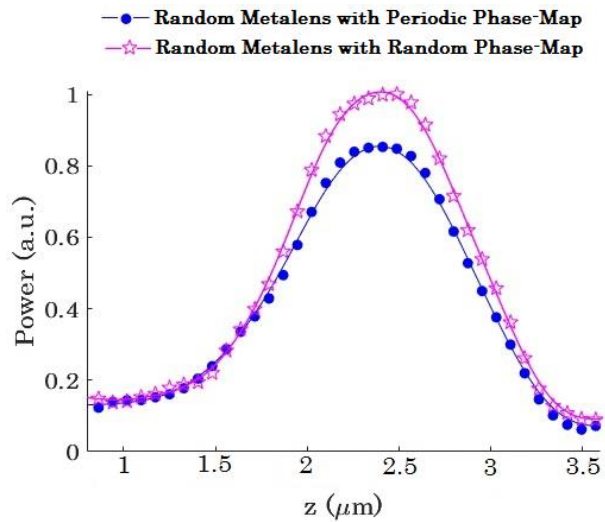

**Fig. 4.** Comparison of the focusing efficiency of random metalens with periodic and random phase-map by plotting the power along the z axis of the random metalenses.

The proposed random-phase map improves the efficiency but further improvements need to take into account other sources of phase errors. For example, the current random phase-map does not fully resolve problems with the periodic phase-map to fully take into account random near-field coupling between elements especially for particles with a smaller radius as they are located far from each other on the phase-map which is not the case on the actual gradient metasurface. Table (1) summarizes possible causes of phase errors in periodic and random meta-devices and the extent to which they are accounted for by periodic and random-phase-maps. As explained before, in the use of periodic phase-map for periodic grid metalenses the implementation neglects the difference in size of neighboring elements which was addressed by the local phase method optimization algorithm [31]. Most causes of errors are not modeled when using periodic phase-map for random metasurfaces, but, with the random phase-map, the number of neighboring particles as well as their angular positions are addressed. The distance between neighboring elements is accounted for in average for particles of large radii but not for particles of small radii.

**Table 1.** Origin of phase errors in periodic and random meta-devices and extent to which they are addressed by the periodic and random phase-maps.

|  | Periodic Phase-Map for Periodic Metalens | Periodic Phase-Map for Random Metalens | Proposed Random Phase-Map for Random Metalens |
|---|---|---|---|
| Number of Neighboring Elements | ✓ | ✗ | ✓ |
| Distance to Neighboring Elements | ✓ | ✗ | -- |
| Size of Neighboring Elements | ✗ | ✗ | ✗ |
| Angular Position of Neighboring Elements | ✓ | ✗ | ✓ |

## Conclusion

We proposed a random phase-map that contains statistical information on random metasurfaces and can be used to efficiently design them. The phase-map provides a simple method to choose the geometry of particles located at arbitrary positions while accounting for near-field coupling between particles and their neighbors in the random metasurface. This approach provides an initial rigorous step of designing efficient random metasurfaces. We have also discussed how random phase-maps can address issues with periodic phase-maps in realizing random meta-devices and reported an increased efficiency using the proposed method. Considering the large degrees of freedom of random media, this work will find applications in information encrypting where random meta-devices with similar performance may encode additional information in the statistics of their particles distribution.